\documentclass[12pt,a4paper]{article}

\usepackage{amsmath}
\usepackage{amssymb}
\usepackage[nosort]{cite} 
\usepackage[hyperref,bulletsep]{collect} 

\makeatletter
\let\old@makecaption=\@makecaption
\def\@makecaption{\small\old@makecaption}
\makeatother

\makeatletter
\@addtoreset{equation}{section}
\makeatother

\makeatletter
\let\old@startsection=\@startsection
\renewcommand{\@startsection}[6]{\old@startsection{#1}{#2}{#3}{#4}{#5}{#6\mathversion{bold}}}
\makeatother

\let\oldPhi=\Phi
\let\oldPsi=\Psi
\let\oldGamma=\Gamma
\let\oldSigma=\Sigma
\renewcommand{\Phi}{\mathnormal{\oldPhi}}
\renewcommand{\Psi}{\mathnormal{\oldPsi}}
\renewcommand{\Gamma}{\mathnormal{\oldGamma}}
\renewcommand{\Sigma}{\mathnormal{\oldSigma}}

\newcommand{\hypref}[2]{\ifx\href\asklfhas #2\else\href{#1}{#2}\fi}

\newcommand{\secref}[1]{Sec.~\ref{#1}}
\newcommand{\appref}[1]{App.~\ref{#1}}
\newcommand{\tabref}[1]{Tab.~\ref{#1}}

\newcommand{\atopfrac}[2]{\genfrac{}{}{0pt}{}{#1}{#2}}
\newcommand{\sfrac}[2]{{\textstyle\frac{#1}{#2}}}
\newcommand{\half}{\sfrac{1}{2}}
\newcommand{\quarter}{\sfrac{1}{4}}

\newcommand{\harm}[1]{h(#1)}

\newcommand{\alg}[1]{\mathfrak{#1}}
\newcommand{\grp}[1]{\mathrm{#1}}
\newcommand{\grSU}{\grp{SU}}

\newcommand{\alSU}{\alg{su}}
\newcommand{\alGL}{\alg{gl}}
\newcommand{\alSL}{\alg{sl}}
\newcommand{\alSO}{\alg{so}}
\newcommand{\alPSL}{\alg{psl}}

\newcommand{\alPSU}{\alg{psu}}

\newcommand{\fldF}{\mathcal{F}}

\newcommand{\superN}{\mathcal{N}}
\newcommand{\gym}{g_{\scriptscriptstyle\mathrm{YM}}}

\newcommand{\Tr}{\mathop{\mathrm{Tr}}}

\newcommand{\indup}[1]{_{\mathrm{#1}}}

\newcommand{\cder}{\mathcal{D}}

\newcommand{\state}[1]{|#1\rangle}%
\newcommand{\vac}{\state{0}}%
\newcommand{\osca}{\mathbf{a}}%
\newcommand{\oscb}{\mathbf{b}}%
\newcommand{\oscc}{\mathbf{c}}%
\newcommand{\oscd}{\mathbf{d}}%
\newcommand{\lrbrk}[1]{\left(#1\right)}
\newcommand{\bigbrk}[1]{\bigl(#1\bigr)}

\newcommand{\lreval}[1]{\left.#1\right|}

\newcommand{\nln}{\nonumber\\}

\newcommand{\eq}{\mathrel{}&=&\mathrel{}}
\newenvironment{myeqnarray}{\arraycolsep0pt\begin{eqnarray}}{\end{eqnarray}\ignorespacesafterend}
\newenvironment{myeqnarray*}{\arraycolsep0pt\begin{eqnarray*}}{\end{eqnarray*}\ignorespacesafterend}

\def\[{\begin{equation}}
\def\]{\end{equation}}
\def\<{\begin{myeqnarray}}
\def\>{\end{myeqnarray}}

\ifx\href\asklfhas\newcommand{\href}[2]{#2}\fi
\newcommand{\arxivno}[1]{\href{http://arxiv.org/abs/#1}{#1}}

\begin{document}


\thispagestyle{empty}
\begin{flushright}\footnotesize
\texttt{\arxivno{hep-th/0307042}}\\
\texttt{AEI 2003-057}
\end{flushright}
\vspace{2cm}

\begin{center}
{\Large\textbf{\mathversion{bold}The $\mathcal{N}=4$ SYM \\Integrable
Super Spin Chain}\par}
\vspace{2cm}

\textsc{Niklas Beisert and Matthias Staudacher}
\vspace{5mm}

\textit{Max-Planck-Institut f\"ur Gravitationsphysik\\
Albert-Einstein-Institut\\
Am M\"uhlenberg 1, D-14476 Golm, Germany}
\vspace{3mm}

\texttt{nbeisert,matthias@aei.mpg.de}\par\vspace{2cm}

\textbf{Abstract}\vspace{7mm}

\begin{minipage}{13.7cm}
Recently it was established that the one-loop planar dilatation 
generator of $\superN=4$ Super Yang-Mills theory may be identified, 
in some restricted cases, with the Hamiltonians of various integrable 
quantum spin chains. In particular Minahan and Zarembo established 
that the restriction to scalar operators leads to an integrable
vector $\alSO(6)$ chain, while recent work in QCD suggested
that restricting to twist operators, containing
mostly covariant derivatives, yields certain integrable Heisenberg XXX
chains with non-compact spin symmetry $\alSL(2)$. Here we unify and generalize
these insights and argue that the \emph{complete} one-loop planar dilatation 
generator of $\superN=4$ is described by an integrable $\alSU(2,2|4)$ 
super spin chain. We also write down various forms
of the associated Bethe ansatz equations, whose solutions are in
one-to-one correspondence with the complete 
set of \emph{all} one-loop planar 
anomalous dimensions in the $\superN=4$ gauge theory. We finally 
speculate on the non-perturbative extension of these integrable 
structures, which appears to involve non-local deformations of the 
conserved charges.

\end{minipage}

\end{center}

\newpage
\setcounter{page}{1}
\setcounter{footnote}{0}


\section{Introduction and Overview}

Of all four-dimensional gauge field theories the one with the maximum
number $\superN=4$ \cite{Gliozzi:1977qd,Brink:1977bc}
of supersymmetries appears to be very special
in many ways. At first sight this model appears to be incredibly 
complicated due to the very large number of fields of different kinds, but
upon closer inspection many hidden simplicities appear. It is
believed that the model is exactly conformally invariant at the quantum 
level \cite{Sohnius:1981sn,Howe:1984sr,Brink:1983pd} and thus 
a non-trivial example (as opposed to free massless field theory) of a 
\emph{four}-dimensional conformal field theory (CFT). Given the
huge successes in understanding CFT's in \emph{two}
dimensions, one might hope that at least some of the aspects
allowing their treatment in $D=2$ might fruitfully reappear in $D=4$.
One of the many intriguing features of two-dimensional CFT's is
that they are intimately connected to integrable 2+0 dimensional
lattice models in statistical mechanics, or, equivalently, 
to 1+1 dimensional quantum spin chains. Integrable models are
even more important for the study of integrable massive deformations
of these CFT's. Thus, an optimistic mind could hope that integrability 
might also play a r\^ole
in the putative simplicity of a theory such as $\superN=4$ Yang-Mills.
Excitingly, a number of recent discoveries lend much support to this idea.
One might wonder about standard, naive no-go theorems that seem to 
suggest that integrability can never exist above $D=2$. These may be
potentially bypassed by the fact that there appears to be a hidden
``two-dimensionality'' in $\grSU(N)$ SYM$_4$ when we look at it at large $N$,
i.e.~when we introduce $\frac{1}{N}$ as an additional coupling 
constant in the theory, aside from the 't~Hooft coupling $\lambda=\gym^2 N$. 

In fact, one might interpret the AdS/CFT correspondence
\cite{Maldacena:1998re,Witten:1998qj,Gubser:1998bc},
believed to be exact in the $\superN=4$ case, 
as one important indication of the validity of this idea:
Again following 't Hooft, the worldsheet of the string is a clear
candidate for explaining the hidden two dimensions. 
Interestingly, first indications that the world sheet theory,
highly non-trivial due to the curved $AdS_5 \times S^5$
background, might be integrable have recently appeared 
\cite{Mandal:2002fs,Bena:2003wd,Vallilo:2003nx} 
(for the simpler but related case of pp-wave backgrounds see also
\cite{Maldacena:2002fy,Russo:2002qj,Bakas:2002kt}).

However, let us get back to the gauge theory and review how 
integrable structures have recently emerged directly, without
resorting to the correspondence. Following the seminal BMN paper
\cite{Berenstein:2002jq} it became clear that, despite much
work on SYM$_4$, not much was known about how to efficiently
calculate the anomalous dimensions of local, non-protected 
conformal operators composed of an arbitrary number of elementary
fields. In particular, in 
\cite{Kristjansen:2002bb,Constable:2002hw,Beisert:2002bb,Constable:2002vq}
effective vertex techniques were developed in order to
efficiently calculate one-loop two-point functions of operators
composed of an arbitrary numbers of scalars.
From these two-point correlators the scaling dimensions could then be 
(indirectly) extracted.
The main focus was on the case where the classical dimension
of the operators tended to infinity, but the technique was
applicable to the finite case as well.
In particular \cite{Beisert:2002bb} contained the full 
$\alSO(6)$ invariant effective vertex which was used in
\cite{Beisert:2002tn} to find the exact finite dimensional
generalizations of the BMN operators.
A significant simplification took place 
when it was realized \cite{Minahan:2002ve,Beisert:2002ff} that the
effective vertex could be directly interpreted as a 
\emph{Hamiltonian} acting on a Hilbert space formed by all possible
scalar operators of fixed classical dimension. Therefore it was no
longer necessary to extract the anomalous dimensions from
a two-point function; instead one could directly go about diagonalizing
the Hamiltonian. This was done in \cite{Beisert:2002ff} for a
Hilbert space containing multi-trace gauge-invariant scalar operators;
a natural split between a planar ``free'' part and a non-planar
trace-splitting and trace-joining ``interaction'' part appeared,
where the interaction is purely cubic with coupling constant $\frac{1}{N}$.
This split turned out to be very reminiscent of similar structures 
appearing in cubic string field theories. In turn, Minahan and Zarembo 
focused on the planar part of the Hamiltonian and noticed the
remarkable fact that it is identical to the one
of an $\alSO(6)$ invariant integrable spin chain. The
integrablity leads, via a hidden Yang-Baxter symmetry, to the 
appearance of an infinite number of charges commuting with the
Hamiltonian. This allows to write down Bethe equations whose
solution furnish the spectrum, i.e.~the planar anomalous dimensions.
For low dimensional operators (i.e.~those containing only a few 
elementary fields) the 
approach is not necessarily easier than direct, brute
force diagonalization of the Hamiltonian (however, it does
lead to a very transparent and short derivation of the finite $J$
BMN operators found in \cite{Beisert:2002tn}).  
Its real power lies in the fact that it allows to obtain 
anomalous dimensions in the limit of a large number of constituent fields.

This power was very recently illustrated in a study of scalar
operators belonging to the $\alSO(6)$ representations
$[J_2,J_1-J_2,J_2]$ with $\Delta_0=J_1+J_2$ 
in the limit where both charges $J_1,J_2$
become large \cite{Beisert:2003xu}.
The ground state energy in this representation is argued
to be related to the minimal energy solution of a string rotating
in two planes in $S^5$, which can be calculated exactly
\cite{Frolov:2003qc,Frolov:2003tu,Frolov:2003xy}.
The result for this energy
is given by inverting certain elliptic functions, it is 
a non-trivial \emph{function} of the parameters $J_1,J_2$,
and agrees on both sides. On the gauge theory side,
we believe it can only be obtained by using the Bethe ansatz.
This is arguably the most subtle dynamical, quantitative test 
of AdS/CFT existing to date.

Integrable spin chains had appeared before in gauge theories through
the pioneering work of Lipatov on high energy scattering in planar QCD
\cite{Lipatov:1994yb}. The model was subsequently identified as
a XXX Heisenberg $\alSL(2)$ spin chain of noncompact spin zero
\cite{Faddeev:1995zg}. More recently, and physically closely
related to the present study, further integrable structures were
discovered by Belitsky, Braun, Derkachov, Korchemsky and Manashov
in the computation of planar one-loop anomalous
dimensions of various types of quasi-partonic operators in
QCD \footnote{While QCD is surely not a conformally invariant quantum field
theory, it still behaves like one as far as one-loop anomalous
dimensions are concerned. 
We thank A.~Belitsky, A.~Gorsky and 
G.~Korchemsky for pointing this out to us.
In one-loop QCD, integrability
is not complete; in particular it requires aligned helicities of
the partonic degrees of freedom.} 
\cite{Braun:1998id,Belitsky:1999qh,Braun:1999te,Belitsky:1999ru,
Belitsky:1999bf,Derkachov:1999ze}. Subsequently, Kotikov and Lipatov 
\cite{Kotikov:2000pm,Kotikov:2001sc,Kotikov:2002ab,Kotikov:2003fb}
applied the corresponding integrable structure, 
originally found in the QCD context,
to the computation of anomalous dimensions of twist-two operators
(i.e.~operators containing two scalar fields and an arbitrary
number of symmetrized, traceless covariant derivatives acting on them) in SYM$_4$. 
The result for the anomalous dimension is given
in terms of harmonic numbers and was
verified by Dolan and Osborn using totally 
different methods \cite{Dolan:2001tt}. 
This result was very useful in (qualitatively) comparing certain string states 
in $AdS_5$ to twist-two operators with a large spin in the gauge theory \cite{Gubser:2002tv}.
Very recently Belitsky, Gorsky and Korchemsky
\cite{Belitsky:2003ys} considered QCD composite light-cone
operators, corresponding to operators of arbitrarily high twist, 
and discussed the hidden integrability of the associated
dilatation operator.

Let us ask the question whether the
integrable structures appearing in the planar gauge theory,
when one investigates the first radiative corrections to 
anomalous dimensions, are accidental or rather more generic and thus
indicative of a deeper underlying structure. The preliminary
evidence from the correspondence with strings \cite{Mandal:2002fs,Bena:2003wd}
certainly points to the second possibility. And indeed a first
study of the two-loop corrections to the integrable Hamiltonian
yields intriguing evidence that these do \emph{not} break the 
integrability property \cite{Beisert:2003tq}. One indirect
way to detect the hidden commuting charges is to look for
unexpected degeneracies in the energy spectrum. We found that
these degeneracies exist in planar SYM$_4$, can be explained at
one loop by the hidden integrability, and are not broken at the
two-loop level. It is important to note that, while the 
one-loop Hamiltonian for scalar fields is exactly integrable
\cite{Minahan:2002ve}, the combined one- and two-loop dilatation
operator is not exactly integrable and requires the inclusion 
of three-loop terms, and so on \cite{Beisert:2003tq}. 
Interpreting these corrections in the spin chain language
one finds that the higher-loop deformations of the charges
are in a certain sense non-local, reminiscent of \cite{Mandal:2002fs,Bena:2003wd}.
The integrable system corresponding to the all-loop planar dilatation
operator of $\superN=4$ Yang-Mills, if it exists, should be
some kind of long-range spin chain and has yet to be identified.
In particular it is unclear how to extend the Bethe ansatz
of \cite{Minahan:2002ve} to include the higher loop effects.

All this suggests that, maybe, planar $\superN=4$ Yang-Mills 
is integrable. If this is the case, the hidden symmetries
should extend the known symmetries of SYM$_4$. 
The full symmetry algebra of SYM$_4$ is neither 
$\alSO(6)$ nor $\alSL(2)$, but the full superconformal algebra
$\alPSU(2,2|4)$.%
\footnote{The conjugation properties of the
algebras are not relevant here, we will always
refer to complex algebras, e.g. $\alSL(4|4)=\alSU(2,2|4)$,
$\alSO(6)=\alSL(4)$, etc.}
If the integrable structures discovered to date
are not accidental, we should expect that the $\alSO(6)$ one-loop results of
Minahan-Zarembo and the $\alSL(2)$ results suggested from
one-loop QCD \cite{Braun:1998id,Belitsky:1999qh,Braun:1999te,Belitsky:1999ru,
Belitsky:1999bf,Derkachov:1999ze,Belitsky:2003ys}
(see also \cite{Kotikov:2000pm,Kotikov:2001sc,Kotikov:2002ab,Kotikov:2003fb})
can be combined and ``lifted'' to a full $\alPSU(2,2|4)$ super spin
chain, as first conjectured%
\footnote{In \cite{Belitsky:2003ys} it was conjectured that the
$\alSL(2)$ and $\alSO(6)$ 
spin chains combine into a 
$\alSO(2,4) \times \alSO(6)=\alSU(2,2) \times \alSU(4)$
chain. As we shall see, we need the full $\alPSU(2,2|4)$ algebra
in order to achieve this ``grand unification''.}
in \cite{Beisert:2003tq}. 
In this paper we will argue that this is indeed the case.
In a companion paper, one of us worked out the complete 
one-loop dilatation operator of $\superN=4$ Yang-Mills theory
\cite{Beisert:2003jj}. That is, the effective vertex (not necessarily
restricted to the planar case) allowing the computation of 
mixing matrices of composite operators containing arbitrary
sequences of elementary fields (scalars, covariant derivatives,
field strengths, and fermions) was derived.
Upon diagonalization of this matrix (which in practice is however 
only possible for states of small classical dimension) its eigensystem
yields the eigenoperators and their anomalous dimensions.
Here we will establish that the restriction of this complete
one-loop dilatation operator to the planar case yields
the promised $\alSU(2,2|4)$~%
\footnote{At the one-loop level
the superconformal algebra $\alPSU(2,2|4)$ may be extended 
by a $\alGL(1)$ charge to the semi-direct product $\alSU(2,2|4)$.}
spin chain. Our arguments
contain some assumptions which should be filled in by
experts on integrability; in particular, while we propose
an R-matrix, we will assume its existence and uniqueness. 
At any rate we then proceed to write down the Bethe ansatz 
equations expected to describe our super spin chain. 
Again, it is not clear to us whether the validity of the ansatz has 
been rigorously proven in the spin chain literature. 
However, by analyzing the first few states
in our superchain we find complete agreement with the gauge
theory results. Therefore we are confident that the details of our
Bethe ansatz are indeed correct. A subtlety, to be discussed below,
is that for superalgebras one can write down \emph{several}
Bethe ans\"atze, which appear to be different and possess different
(pseudo)vacuum states. They nevertheless yield an identical
spectrum. We will discuss two specific ans\"atze which are both natural
from different points of view.

The outline of this paper is as follows. 
We begin with a short review/preview on Bethe ansatz equations,
and how they are determined by the Dynkin diagram of the symmetry
algebra of the chain in conjunction with the highest weight representation 
of the degrees of freedom distributed along the chain.
We will also introduce some useful, compact notation for writing
the Bethe equations. We explain how to employ the latter 
in order to derive anomalous dimensions in 
$\superN=4$ Yang-Mills. Next we present our
arguments, via the existence of a unique R-matrix,
why one expects the complete one-loop dilatation operator
of \cite{Beisert:2003jj} to become integrable in the large $N$ limit.
The procedure naturally incorporates the $\alSL(2)$
(twist operators) and the $\alSO(6)$
(Minahan-Zarembo) integrable structures. 
For super chains we can write seemingly different sets of Bethe
ansatz equations. 
We will pick two examples: The first (which we call ``Beauty'') 
does not correspond to the standard way of choosing the
fermionic root(s) of the Dynkin diagram. However, this
formulation is particularly useful since the vacuum states of
the chain correspond to the half-BPS states of SYM$_4$.
To illustrate the freedom of choosing various forms of the Bethe
equations we then present a second version (``the Beast'')
which employs the standard ``distinguished'' choice of one
fermionic root in the middle of the diagram. As we will 
see, this shifts the vacuum to a different, high energy state
containing only field strengths. It obscures certain well known
results (e.g.~it becomes hard to see that half-BPS states have
vanishing energy) but instead leads to some further interesting
spectroscopic results. We end with an outlook on
interesting applications and extensions of this work. 

\section{Anomalous dimensions, (super) spin chains, and Bethe ans\"atze}
\label{sec:TechIntro}

Let us review how anomalous dimensions in YM$_4$ are computed
by employing the algebraic Bethe ansatz, once it has been 
established that the anomalous part of the dilatation operator 
is identical to the Hamiltonian of
an integrable quantum spin chain \cite{Minahan:2002ve}. 
Let us explain this in the simplest case of an $\alSL(2)$ chain,
the so-called XXX$_{s/2}$ Heisenberg chain.  
(For a very pedagogical introduction, see \cite{Faddeev:1996iy}).
The proposal is
that the energy eigenvalues $E$ of the Hamiltonian are proportional
to the anomalous dimensions $\delta \Delta$ of conformal scaling operators:
\[\label{eq:dim}
\delta \Delta=\frac{\gym^2 N}{8\pi^2}\,E.
\]
For spin $\frac{s}{2}=\frac{1}{2}$ these scaling operators are, in the planar
limit, linear combinations of a single trace of sequences of 
two of the three complex scalars of SYM$_4$, say $Z$ and $\phi$:
\[\label{eq:updown}
\Tr \bigbrk{ Z Z Z \phi Z \phi \phi ... }
\]
The single trace operator is then interpreted as a chain of
spins, where $Z$ and $\phi$ represent, respectively, up and
down spins. The total number $L$ of complex scalars inside the trace
corresponds to the length of the chain.
Each eigenstate (alias scaling operator) of the Hamiltonian is 
uniquely characterized by a set of Bethe roots $u_j$, $j=1,\ldots,n$,
and the energy $E$ of the state is given by  
\[\label{eq:energy}
E=\pm  \sum_{j=1}^{n} \frac{s}{u_j^2+\sfrac{1}{4}s^2}.
\]
(where for spin $\frac{s}{2}=\frac{1}{2}$ we should pick the plus sign
in front of the sum).
The Bethe roots are found by solving the Bethe equations
for $j=1,\ldots,n$
\[\label{eq:Bethesl2}
\lrbrk{\frac{u_j+\sfrac{i}{2} s}{u_j-\sfrac{i}{2} s}}^L=
\prod_{\textstyle\atopfrac{k=1}{k\neq j}}^n\frac{u_j-u_k+i}{u_j-u_k-i}.
\]
The vacuum state of energy $E=0$ is the half-BPS state
\[\label{vac}
\vac=\Tr Z^L,
\]
i.e.~it is interpreted as the ferromagnetic ground state
(all spins are up, no Bethe roots) of the chain. 
The excitation number $n$, giving the total number of roots, 
counts the number of $\phi$'s or
down spins along the chain;
it is naturally bounded by $n\leq L$. 
There is an additional constraint on the Bethe roots:
\[\label{eq:momentum}
1=\prod_{j=1}^n\frac{u_j+\sfrac{i}{2} s}{u_j-\sfrac{i}{2} s}.
\]
For the spin chain, it means that we have periodic boundary conditions
and we are only looking for zero-momentum states. In the
gauge theory interpretation \eqref{eq:updown} it expresses the
cyclicity of the trace. 

The second simplest example concerns twist operators,
closely related to quasi-partonic operators in one-loop QCD
\cite{Braun:1998id,Belitsky:1999qh,Braun:1999te,Belitsky:1999ru,
Belitsky:1999bf,Derkachov:1999ze,Kotikov:2000pm,Kotikov:2001sc,
Kotikov:2002ab,Kotikov:2003fb,Belitsky:2003ys}.
Just as the fields in \eqref{eq:updown} form a closed subsector 
(i.e.~they do not mix with any other types of fields)
the following operators form another such subsector 
(see also \cite{Beisert:2003jj} for further details):
\[\label{eq:baby}
\Tr \bigbrk{ ({\cal D}^{m_1}Z) ({\cal D}^{m_2}Z) \ldots
({\cal D}^{m_L}Z) }
\]
where ${\cal D}$ is the ``light-cone'' covariant derivative 
${\cal D}_{1+i 2}$ (as in \cite{Belitsky:2003ys}). This is
still a spin chain of length (twist) $L$, with non-compact spin 
$\frac{s}{2}=-\frac{1}{2}$,
as will be proven in the next chapter.
Here the spins at each lattice site $k$ may take any value
$m_k=0,1,2,\ldots$, as we have an infinite $s=-1$
representation of $\alSL(2)$. 
Also the total excitation number $n=\sum m_k$ is not bounded
as in the above example.
The vacuum is still
$\Tr Z^L$. 
Note that in this example the length of the chain
is \emph{not} equal to the classical dimension $L+n$ of the operator.
Again, the anomalous dimensions of the scaling operators formed
from the states \eqref{eq:baby} are given via 
eqs.~\eqref{eq:dim},\eqref{eq:energy},\eqref{eq:Bethesl2},\eqref{eq:momentum}\,!
(Here we have to choose the negative sign in the energy formula 
\eqref{eq:energy}.) 

In the above example the algebra is $\alSL(2)$ and thus of rank one.
There is a beautiful extension of the Bethe equations to any algebra
and any representation of the elementary constituents of the chain
at each lattice site 
\cite{Reshetikhin:1983vw,Reshetikhin:1985vd,Ogievetsky:1986hu}.
This general form easily extends to the case of super algebras
as well, see \cite{Saleur:1999cx} and references therein.
This is what we need when we are looking for
a spin chain describing SYM$_4$ at one loop, where we
expect Bethe equations for the super algebra $\alSL(4|4)$.
This general equation may be written in the following compact
notation, based on knowing the Dynkin diagram of the algebra.
The Dynkin diagram of $\alSL(4|4)$ contains seven dots corresponding
to a choice of seven simple roots.
Consider a total of $n$ excitations.
For each of the corresponding Bethe roots $u_i$, $i=1,\ldots,n$,
we specify which of the seven simple roots is excited by
$k_j=1,\ldots,7$.
The Bethe equations for $j=1,\ldots,n$
can then be written in the compact form 
\[\label{eq:BeautyBethe}
\lrbrk{\frac{u_j+\sfrac{i}{2}V_{k_j}}{u_j-\sfrac{i}{2}V_{k_j}}}^L=
\prod_{\textstyle\atopfrac{l=1}{l\neq j}}^n\frac{u_j-u_l+\sfrac{i}{2}M_{k_j,k_l}}{u_j-u_l-\sfrac{i}{2}M_{k_j,k_l}}.
\]
Here, $M_{kl}$ is the Cartan matrix of the algebra and $V_k$ 
are the Dynkin labels of the spin representation.
Furthermore, we still consider a cyclic spin chain with zero
total momentum. This gives the additional constraint
\[\label{eq:BeautyMomentum}
1=\prod_{j=1}^n\frac{u_j+\sfrac{i}{2}V_{k_j}}{u_j-\sfrac{i}{2}V_{k_j}}.
\]
The energy of a configuration of roots that
satisfies the Bethe equations and constraint is now given by
\footnote{The Bethe equations determine the energy 
only up to scale and a shift $c L$ where c is a constant to
be fixed 
such that the energy corresponds
to planar anomalous dimensions of $\superN=4$ SYM
via \eqref{eq:dim}.}
\[\label{eq:NotBeautyEnergy}
E=c L \pm \sum_{j=1}^n\lrbrk{\frac{i}{u_j+\sfrac{i}{2}V_{k_j}}-\frac{i}{u_j-\sfrac{i}{2}V_{k_j}}}.
\]
It is easily seen that restricting these equations to the Dynkin
diagram of the algebra $\alSO(6)$ reproduces the Bethe equations of
\cite{Minahan:2002ve}. It will turn out, see below, that these 
general equations,
which are well known in the literature on integrable spin chains,
indeed solve the entire problem of computing planar anomalous
dimensions in SYM$_4$, once we (\emph{i}) identify the 
correct representations of the fundamental fields on the lattice
sites, and (\emph{ii}) after resolving certain subtleties concerning
Dynkin diagrams for superalgebras. However, let us first 
show that we expect planar one-loop SYM$_4$ to be 
completely integrable indeed.

\section{The R-Matrix}
\label{sec:RMatrix}

In this section we derive the R-matrix 
for the integrable spin chain 
considered in this work. 
For this we make use of a special subsector 
of the spin chain
with residual $\alSL(2)$ symmetry
and show how to 
lift the universal $\alSL(2)$ R-matrix 
to an $\alSL(4|4)$ invariant R-matrix.
The derived Hamiltonian is shown
to agree with the complete one-loop planar dilatation generator 
of $\superN=4$ SYM, thus proving the integrability of the latter.

\paragraph{The planar one-loop dilatation operator.}

The complete one-loop planar dilatation generator 
of $\superN=4$ SYM 
acting on two fields is given by \cite{Beisert:2003jj}
\[\label{eq:SL44Ham}
H_{12}=2h(J_{12}):=
\sum_{j=0}^\infty
2 \harm{j}\, P_{12,j}.
\]
Here, $h(j)$ is the $j$-th harmonic number defined
alternatively by a series or by the digamma function
$\Psi(x)=\partial\log\Gamma(x)/\partial x$
\[\label{eq:SL44Harmonic}
h(n)=\sum_{k=1}^n\frac{1}{k}=\Psi(n+1)-\Psi(1).
\]
The operator $P_{12,j}$ projects 
the spins at sites $1,2$ to the module $V_j$.
Each spin belongs to the `fundamental' module 
$V\indup{F}$, consequently the modules $V_j$ arise 
in the tensor product of two fundamental modules 
\[\label{eq:SL44Tensor}
V\indup{F}\times V\indup{F}=
\sum_{j=0}^\infty V_{j}.\]
The primary weights of these modules are given by
(see \eqref{eq:BeautyWeight} for details)
\<\label{eq:SL44Weights}
w\indup{F}\eq[1;0,0;0,1,0;0,1],\nln
w_0\eq [2;0,0;0,2,0;0,2],\nln
w_1\eq [2;0,0;1,0,1;0,2],\nln
w_{j}\eq [j;j-2,j-2;0,0,0;0,2].
\>

\paragraph{The $\alSL(2)$ subsector.}

A crucial point in the derivation of the
full Hamiltonian density \cite{Beisert:2003jj}
was the restriction to a subsector.
This subsector is obtained by restricting the allowed weights of states
to 
\[
\label{eq:SL2Weight}
w= [L+s;s,s;0,L,0;0,L].\]
For the construction of such states we must restrict all
spins to the form 
\[\label{eq:SL2Spin}
z^n:=\frac{1}{n!}\,\cder^n Z, \qquad\mbox{with}\quad \cder:=\cder_1+i\cder_2,\quad Z:=\Phi_5+i\Phi_6.\]
The residual symmetry algebra within the subsector is
$\alSL(2)$ which can be represented by
the generators
\[\label{eq:SL2Alg}
J'_-= \partial,
\quad
J'_3= \half+z\partial,
\quad
J'_+= z+z^2\partial
\]
where $\partial$ is the derivative with respect to $z$.
It was then shown that the modules $V\indup{F},V_j$ of the full theory 
correspond to the $\alSL(2)$ modules $V'\indup{F},V'_j$ with highest weights
\[\label{eq:SL2Weights}
w'\indup{F}=[-1],\quad w'_j=[-2-2j].
\]
Here, the weights are described by \emph{twice} the spin.
The Hamiltonian density \eqref{eq:SL44Ham} restricts within this subsector to 
\[\label{eq:SL2Ham}
H'_{12}=2h(J'_{12}):=
\sum_{j=0}^\infty
2 \harm{j}\, P'_{12,j},
\]
where $P'_{12,j}$ projects states of the 
tensor product $V'\indup{F}\times V'\indup{F}$ to
the module $V'_j$.
Interestingly, this is the Hamiltonian of
the XXX$_{-1/2}$ Heisenberg spin chain.
In other words it is an integrable spin chain 
with unbroken $\alSL(2)$ symmetry, where
each site transforms in a spin $w'\indup{F}/2=-\sfrac{1}{2}$ representation.

\paragraph{Integrability within the $\alSL(2)$ subsector.}

To prove this statement we make use of the universal 
R-matrix of $\alSL(2)$ spin chains. This $\alSL(2)$ invariant 
operator can be decomposed into its irreducible components 
corresponding to the modules $V'_j$
\[\label{eq:SL2Rop}
R'_{12}(u)=\sum_{j=0}^\infty R'_j(u)\, P'_{12,j}.
\]
The eigenvalues $R'_j(u)$ of the $\alSL(2)$ universal R-matrix 
were determined in \cite{Kulish:1981gi}. In a 
spin $w'_j/2=-1-j$ representation the eigenvalue is 
\[\label{eq:SL2Rval}
R'_{j}(u)=(-1)^{j+1}\,\frac{\Gamma(-j-cu)}{\Gamma(-j+cu)}\,f(cu).
\]
The arbitrary function $f(u)$ and normalization constant $c$ 
reflect the trivial symmetries of the 
Yang-Baxter equation.
The induced eigenvalues of the Hamiltonian density are obtained 
as the logarithmic derivative of $R'_{j}(u)$ at $u=0$ times 
the permutation operator $P_{12}$
\[\label{eq:SL2HamFromR}
H_{12}\, V'_j=P_{12}\,\lreval{\frac{\partial \log R'_{j}(u)}{\partial u}}_{u=0}\, V'_j.\]
We note that for even (odd) $j$ the composite module
$V'_j$ is a (anti)symmetric combination 
of two $V'\indup{F}$, consequently the permutation acts as
\[\label{eq:SL2Perm}
P_{12}\,V'_j=(-1)^j\,V'_j.
\]
We choose the function and constant to be
\[\label{eq:SL2Norm}
f(cu)=-\frac{\Gamma(-cu)}{\Gamma(cu)},\quad c=1.
\]
Using the fact that $j$ is exactly integer
we find the eigenvalues of the
Hamiltonian density
\[\label{eq:SL2HamHarm}
\lreval{(-1)^j\,\frac{\partial \log R'_{j}(u)}{\partial u}}_{u=0}
=2\Psi(j+1)-2\Psi(1)=2\harm{j}.\]
This proves that the Hamiltonian density
\eqref{eq:SL2Ham} is integrable.

\paragraph{Integrability of the full $\alSL(4|4)$ R-matrix.}

To derive an R-matrix for the full $\alSL(4|4)$ spin chain 
we will assume that for given representations of the
symmetry algebra there exists a unique R-matrix 
which satisfies the Yang-Baxter equation
(modulo the symmetries of the YBE).
We are not aware whether this claim
\cite{Kulish:1981gi} has been proven.
Let $R_{12}$ be this R-matrix for the
$\alSL(4|4)$ integrable spin chain.
The R-matrix is an invariant operator,
thus it can be reduced to its irreducible components 
corresponding to the modules $V_j$
\[\label{eq:SL44Rop}
R_{12}(u)=\sum_{j=0}^\infty R_{j}(u)\, P_{12,j}.
\]
The restriction $R'$ of the R-matrix to the $\alSL(2)$ sector 
must also satisfy the Yang-Baxter equation.
The unique solution for the eigenvalues of $R'$ is \eqref{eq:SL2Rval}.
Due to the one-to-one correspondence of modules $V_j$ and $V'_j$, 
\[\label{eq:SL44SL2}
V'_j\subset V_j
\]
the eigenvalues of the 
unique $\alSL(4|4)$ R-matrix \eqref{eq:SL44Rop} must be
\[\label{eq:SL44Rval}
R_{j}(u)=
(-1)^{j}\, \frac{\Gamma(-j-cu)}{\Gamma(-j+cu)}\, f(cu).\]
As in \eqref{eq:SL2HamHarm} this R-matrix yields $H_{12}=2\harm{J_{12}}$
in agreement with \eqref{eq:SL44Ham}.
This in turn shows that the planar one-loop
dilatation generator of $\superN=4$ is integrable.
Note however, that this proof is based on the assumption
of the existence of a unique R-matrix.

\paragraph{The Minahan and Zarembo chain.}

As an application 
let us investigate the restriction of the R-matrix 
to the $\alSO(6)$ subsector investigated by
Minahan and Zarembo \cite{Minahan:2002ve}
\[\label{eq:MZWeights}
w= [L;0,0;q_1,p,q_2;0,L],\]
i.e. where the spins are given by the $\superN=4$ SYM
scalars $\Phi_m$.
Two scalars $\Phi_p,\Phi_q$ can transform 
in three different irreducible representations of $\alSO(6)$,
symmetric-traceless, antisymmetric and singlet.
These correspond to the modules $V_0,V_1,V_2$, respectively.

We set the normalization function and constant to 
\[\label{eq:MZNorm}
f(u)=\frac{\Gamma(3+cu)}{2\Gamma(1-cu)},\quad c=1.
\]
and compute the eigenvalues of the R-matrix \eqref{eq:SL44Rval}
for $V_0,V_1,V_2$
\<\label{eq:MZEigen}
R_{12,0}(u)\eq\half (u+1)(u+2),\nln
R_{12,1}(u)\eq\half (u-1)(u+2),\nln
R_{12,2}(u)\eq\half (u-1)(u-2).
\>
We note the projectors to $V_0,V_1,V_2$ in the scalar sector
\<\label{eq:MZProj}
P_{12,0}\eq \half+\half P_{12}-\sfrac{1}{6}K_{12},
\nln
P_{12,1}\eq\half-\half P_{12},
\nln
P_{12,2}\eq\sfrac{1}{6}K_{12},
\>
where $K_{12}$ is the $\alSO(6)$-trace operator
defined by
$K_{12}\Phi_{1,m}\Phi_{2,n}=\delta_{mn}\Phi_{1,p}\Phi_{2,p}$.
Assembling \eqref{eq:MZEigen} and \eqref{eq:MZProj} we
obtain the R-matrix operator \eqref{eq:SL44Rop}
\[\label{eq:MZR}
R_{12}(u)=\half (u+2)P_{12}+\half u(u+2)-\half u K_{12}.
\]
This is precisely the R-matrix of the Minahan and Zarembo
chain \cite{Minahan:2002ve} which has previously been found
by Reshetikhin \cite{Reshetikhin:1983vw,Reshetikhin:1985vd}.

\section{Beauty \ldots}
\label{sec:Beauty}

In the last section we have established that
the planar one-loop dilatation operator 
of $\superN=4$ YM$_4$ is integrable. We therefore expect the general
Bethe ansatz equations \eqref{eq:BeautyBethe} to hold. 
However, for them to be useful, we still need to specify the Dynkin 
labels, the Cartan matrix and precise form of the energy
\eqref{eq:NotBeautyEnergy}.
Furthermore, we will perform a check of the validity of this 
$\alSL(4|4)$ Bethe ansatz
which goes beyond the $\alSO(6)$ subsector. 

\paragraph{Representations.}

First, we need to specify the Cartan matrix,
determined by the Dynkin diagram, and the Dynkin labels of the
spin representation corresponding to the module $V\indup{F}$.
For classical semi-simple Lie algebra the Dynkin diagram is unique.
In the case of superalgebras, however, there is some freedom to distribute
the simple fermionic roots. 
In the context of $\superN=4$ SYM 
one particular choice of Dynkin diagram turns
out to be very useful \cite{Dobrev:1987qz}:%
\footnote{We thank V. Dobrev for this hint, and for informing us about
reference \cite{Dobrev:1987qz}.}
\[\label{eq:BeautyDynkin}
\begin{minipage}{260pt}
\setlength{\unitlength}{1pt}%
\small\thicklines%
\begin{picture}(260,35)(-10,-10)
\put(  0,00){\circle{15}}%
\put(  0,10){\makebox(0,0)[b]{}}%
\put(  7,00){\line(1,0){26}}%
\put( 40,00){\circle{15}}%
\put( 40,15){\makebox(0,0)[b]{}}%
\put( 47,00){\line(1,0){26}}%
\put( 80,00){\circle{15}}%
\put( 80,15){\makebox(0,0)[b]{}}%
\put( 87,00){\line(1,0){26}}%
\put(120,00){\circle{15}}%
\put(120,15){\makebox(0,0)[b]{+1}}%
\put(127,00){\line(1,0){26}}%
\put(160,00){\circle{15}}%
\put(160,15){\makebox(0,0)[b]{}}%
\put(167,00){\line(1,0){26}}%
\put(200,00){\circle{15}}%
\put(200,15){\makebox(0,0)[b]{}}%
\put(207,00){\line(1,0){26}}%
\put(240,00){\circle{15}}%
\put(240,15){\makebox(0,0)[b]{}}%
\put( 35,-5){\line(1, 1){10}}%
\put( 35, 5){\line(1,-1){10}}%
\put(195,-5){\line(1, 1){10}}%
\put(195, 5){\line(1,-1){10}}%
\end{picture}
\end{minipage}
\]
On top of the Dynkin diagram 
we have indicated the
Dynkin labels of the spin representation.
We write the Cartan matrix corresponding to this choice of Dynkin diagram
and the representation vector as%
\footnote{In fact, the Cartan matrix is obtained 
from this by inverting some lines.
The Bethe equations are invariant under the
inversion and it is slightly more
convenient to work with a symmetric matrix $M$.}
\[\label{eq:BeautyMatrix}
M=\left(\begin{array}{c|c|ccc|c|c}
-2&+1&  &  &  &  &   \\\hline   
+1&  &-1&  &  &  &   \\\hline
  &-1&+2&-1&  &  &   \\
  &  &-1&+2&-1&  &   \\
  &  &  &-1&+2&-1&   \\\hline
  &  &  &  &-1&  &+1 \\\hline
  &  &  &  &  &+1&-2
\end{array}\right),\qquad
V=\left(\begin{array}{r}
0\\\hline0\\\hline0\\1\\0\\\hline0\\\hline0
\end{array}\right).
\]
The energy corresponding to a solution to the Bethe equations is 
\[\label{eq:BeautyEnergy}
E=\sum_{j=1}^n\lrbrk{\frac{i}{u_j+\sfrac{i}{2}V_{k_j}}-\frac{i}{u_j-\sfrac{i}{2}V_{k_j}}}.
\]
%

\paragraph{Excitation numbers.}
Finally, we need to obtain the number of excitations $n_k$,
$k=1,\ldots,7$ of the individual simple roots 
for a state with a given weight
\[\label{eq:BeautyWeight}
w=[\Delta_0;s_1,s_2;q_1,p,q_2;B,L].
\]
A weight of (the classical) $\alSL(4|4)$ is described by various labels.
The (bare) dimension is indicated by $\Delta_0$.
The $\alSL(4)$ and $\alSL(2)\times \alSL(2)$ Dynkin labels 
are given by $[q_1,p,q_2]$ and $[s_1,s_2]$, respectively, 
where $s_1,s_2$ equal \emph{twice} the spin.
Furthermore, $B$ describes the $\alGL(1)$ hypercharge or chirality
in the semi-direct product
$\alSL(4|4)=\alGL(1)\ltimes \alPSL(4|4)$.
Finally, $L$, which is not a label of $\alSL(4|4)$,
gives the length or number of spins.
In $\superN=4$ SYM the chirality and length of
an operator are not good quantum numbers,
they are broken at higher loops by the Konishi anomaly. 
Nevertheless, at one-loop order they are conserved and
their leading order values can be used to describe a state.

This is most easily seen in the oscillator 
picture in \cite{Beisert:2003jj}
using the physical vacuum $\state{Z}$.
We write down the action 
of the generators
corresponding to the simple roots
in terms of creation and annihilation operators
\[\label{eq:BeautyRoots}
\begin{minipage}{260pt}
\setlength{\unitlength}{1pt}%
\small\thicklines%
\begin{picture}(260,55)(-10,-30)
\put(  0,00){\circle{15}}%
\put(  0,15){\makebox(0,0)[b]{$n_1$}}%
\put(  0,-15){\makebox(0,0)[t]{$\osca^\dagger_2 \osca^1$}}%
\put(  7,00){\line(1,0){26}}%
\put( 40,00){\circle{15}}%
\put( 40,15){\makebox(0,0)[b]{$n_2$}}%
\put( 40,-15){\makebox(0,0)[t]{$\osca^\dagger_1 \oscc^1$}}%
\put( 47,00){\line(1,0){26}}%
\put( 80,00){\circle{15}}%
\put( 80,15){\makebox(0,0)[b]{$n_3$}}%
\put( 80,-15){\makebox(0,0)[t]{$\oscc^\dagger_1 \oscc^2$}}%
\put( 87,00){\line(1,0){26}}%
\put(120,00){\circle{15}}%
\put(120,15){\makebox(0,0)[b]{$n_4$}}%
\put(120,-15){\makebox(0,0)[t]{$\oscc^\dagger_2 \oscd^\dagger_1$}}%
\put(127,00){\line(1,0){26}}%
\put(160,00){\circle{15}}%
\put(160,15){\makebox(0,0)[b]{$n_5$}}%
\put(160,-15){\makebox(0,0)[t]{$\oscd^\dagger_2 \oscd^1$}}%
\put(167,00){\line(1,0){26}}%
\put(200,00){\circle{15}}%
\put(200,15){\makebox(0,0)[b]{$n_6$}}%
\put(200,-15){\makebox(0,0)[t]{$\oscb^\dagger_1 \oscd^2$}}%
\put(207,00){\line(1,0){26}}%
\put(240,00){\circle{15}}%
\put(240,15){\makebox(0,0)[b]{$n_7$}}%
\put(240,-15){\makebox(0,0)[t]{$\oscb^\dagger_2 \oscb^1$}}%
\put( 35,-5){\line(1, 1){10}}%
\put( 35, 5){\line(1,-1){10}}%
\put(195,-5){\line(1, 1){10}}%
\put(195, 5){\line(1,-1){10}}%
\end{picture}
\end{minipage}
\]
It is now clear that 
$n_1=n_{\osca^\dagger_2}$,
$n_2=n_{\osca^\dagger_1}+n_{\osca^\dagger_2}$
and so on. Using the formulas in \cite{Beisert:2003jj}
we write down the corresponding excitation
numbers of the simple roots
\[\label{eq:BeautyExcite}
n_k=\left(\begin{array}{l}
\half\Delta_0-\sfrac{1}{2}(L-B)-\half s_1\\
\phantom{\half}\Delta_0-\phantom{\half}(L-B)\\
\phantom{\half}\Delta_0-\half(L-B)-\half p-\sfrac{3}{4}q_1-\sfrac{1}{4}q_2\\
\phantom{\half}\Delta_0\phantom{\mathord{}-\sfrac{0}{2}(L-B)}-\phantom{\half}p-\half q_1-\half q_2\\
\phantom{\half}\Delta_0-\half(L+B)-\half p-\sfrac{1}{4}q_1-\sfrac{3}{4}q_2\\
\phantom{\half}\Delta_0-\phantom{\half}(L+B)\\
\half\Delta_0-\half(L+B)-\half s_2
\end{array}\right).\]
Not all excitations of the simple roots correspond to
physical states. Obviously, the excitation numbers of the
oscillators must be non-negative, this gives the bounds
\footnote{
    Superconformal primaries reside in the fundamental Weyl chamber 
    defined by the bounds
    $-2n_1+n_2>-1$, 
    $n_2-2n_3+n_4>-1$, 
    $n_3-2n_4+n_5+L>-1$, 
    $n_4-2n_5+n_6>-1$, 
    $n_6-2n_7>-1$.
    Together with \eqref{eq:BeautyBounds} this implies, among other relations, 
    \eqref{eq:BeautyBounds2}.
    Solutions of the Bethe equations outside the fundamental 
    domain apparently correspond to mirror images of primary states 
    due to reflections at the chamber boundaries (We thank
    J.~Minahan and K.~Zarembo for this insight).
}
\[\label{eq:BeautyBounds}
0\leq n_1\leq n_2\leq n_3\leq 
n_4\geq n_5\geq n_6\geq n_7\geq 0.
\]
Furthermore, each fermionic oscillator cannot be excited more than
once, this gives the bounds
\[\label{eq:BeautyBounds2}
n_2+2L\geq n_3+L\geq n_4\leq n_5+L\leq n_6+2L.\]

Certainly, we should obtain the $\alSO(6)=\alSL(4)$ subsector 
studied by Minahan and Zarembo \cite{Minahan:2002ve}
when we remove the outer four simple roots from the
Dynkin diagram in \eqref{eq:BeautyDynkin}.
When we restrict to the states \eqref{eq:MZWeights} of this 
subsector the number of excitations \eqref{eq:BeautyExcite} of
the outer four roots is trivially zero. They become
irrelevant for the Bethe ansatz and can be discarded.
Thus all solutions to the $\alSO(6)$ Bethe equations
are also solutions to the $\alSL(4|4)$ Bethe equations.
What is more, we can apply this Bethe ansatz to a 
wider range of operators, in fact, to \emph{all}
single-trace operators of $\superN=4$ SYM.

\paragraph{Multiplet splitting.}

Now we can write down and try to solve the Bethe equations 
for any state in $\superN=4$ SYM. 
Note, however, that the Bethe equations need to be solved 
only for highest weight states. 
All descendants of a highest weight state are obtained by
adding Bethe roots at infinity, $u_i=\infty$. 
In other words, the solutions to the Bethe equations
corresponding to highest weight states
are distinguished in that they have no roots $u_i$ at infinity.
Nevertheless, there is one subtlety related to this point
which can be used to our advantage.
Namely this is multiplet splitting at the unitarity bounds
\cite{Dobrev:1985qv,Andrianopoli:1999vr}.
We assume that the spin chain of $L$ sites transforms in the
tensor product of $L$ spin representations. 
The corrections $\delta \Delta$ to the scaling dimension
induced by the Hamiltonian $H$ are not included in this picture.
Thus, in terms of the spin chain only the classical $\alSL(4|4)$ algebra applies
where the scaling dimension is exactly $\Delta_0$.
Long multiplets of the interacting $\alSL(4|4)$ algebra
close to one or both of the unitarity bounds
\<\label{eq:BeautyShorten}
\mathrm{I}:&\quad& \Delta_0=2+s_1+p+\sfrac{3}{2}q_1+\sfrac{1}{2}q_2,\nln
\mathrm{II}:&\quad& \Delta_0=2+s_2+p+\sfrac{1}{2}q_1+\sfrac{3}{2}q_2.
\>
split up into several semi-short or BPS multiplets.
The weights of the additional primary states are offset by
\<\label{eq:BeautyOffsets}
\delta w\indup{I}\eq
\left\{\begin{array}{lcl}
[+0.5;-1,\phantom{+}0;+1,\phantom{+}0,\phantom{+}0;-0.5,+1],&\quad&\mbox{for }s_1>0,\\{}
[+1.0;\phantom{+}0,\phantom{+}0;+2,\phantom{+}0,\phantom{+}0;\phantom{+}0.0,+1],&&\mbox{for }s_1=0,
\end{array}\right.
\nln
\delta w\indup{II}\eq
\left\{\begin{array}{lcl}
[+0.5;\phantom{+}0,-1;\phantom{+}0,\phantom{+}0,+1;+0.5,+1],&\quad&\mbox{for }s_2>0,\\{}
[+1.0;\phantom{+}0,\phantom{+}0;\phantom{+}0,\phantom{+}0,+2;\phantom{+}0.0,+1],&&\mbox{for }s_2=0.
\end{array}\right.
\>
The unitarity bounds can also be expressed in terms
of excitations of simple roots, we find 
\<\label{eq:BeautyShorten2}
\mathrm{I}:&\quad& n_1+n_3=n_2+1,\nln
\mathrm{II}:&\quad& n_7+n_5=n_6+1.
\>
The corresponding offsets translate into
\<\label{eq:BeautyOffsets2}
\delta n\indup{I}\eq
\left\{\begin{array}{lcl}
(\phantom{+}0,-1,-1,\phantom{+}0,\phantom{+}0,\phantom{+}0,\phantom{+}0),&\quad&\mbox{for }n_2>n_1,\\{}
(\phantom{+}0,\phantom{+}0,-1,\phantom{+}0,\phantom{+}0,\phantom{+}0,\phantom{+}0),&\quad&\mbox{for }n_2=n_1,
\end{array}\right.
\nln
\delta n\indup{II}\eq
\left\{\begin{array}{lcl}
(\phantom{+}0,\phantom{+}0,\phantom{+}0,\phantom{+}0,-1,-1,\phantom{+}0),&\quad&\mbox{for }n_6>n_7,\\{}
(\phantom{+}0,\phantom{+}0,\phantom{+}0,\phantom{+}0,-1,\phantom{+}0,\phantom{+}0),&\quad&\mbox{for }n_6=n_7,
\end{array}\right.
\>
together with an increase of the length $L$ by one. 
We thus see that in the case of multiplet shortening
the primaries of higher submultiplets have less excitations. 
In a calculation this may reduce the complexity 
of the Bethe equations somewhat as we shall see in an example below.

Multiplet splitting is an extremely interesting
issue from the point of view of integrability.
Let us consider some operator acting on a
spin chain. Assume the operator
is invariant under the classical 
algebra $\alSL(4|4)$. In the most 
general case, this operator can assign 
a different value to all irreducible multiplets
of states.
In particular this is so for the submultiplets 
of a long multiplet at the unitarity bound. 
Now, if we impose integrability on the operator
all these submultiplets become degenerate.%
\footnote{This is because the unique 
integrable operator is equivalent
to the one-loop planar correction to the 
dilatation operator of $\superN=4$ SYM.
In $\superN=4$ SYM the
submultiplets must rejoin into a long multiplet
which would be inconsistent if their
one-loop anomalous dimensions were different.}
A priori, this seems like a miracle. 
Why should integrability imply \emph{this} degeneracy?
It almost seems as if integrability selects that scalar operator
which is suitable as a consistent deformation
of the dilatation generator!
Then, clearly the miracle would turn into
the condition for integrability.
If this can be made sense of, maybe it also
helps to understand higher loop corrections 
in the light of integrability \cite{Beisert:2003tq}.

\paragraph{Example.}

Let us apply the Bethe ansatz to the twist-two operator with 
primary weight
\[\label{eq:BeautyTwist}
w=[4;2,2;0,0,0;0,2].
\]
Using \eqref{eq:BeautyExcite}
we find the excitation numbers and length
\[\label{eq:BeautyTwistExcite}
n_{0,k}=(0,2,3,4,3,2,0),\quad L_0=2.
\]
This weight is on both unitarity bounds, \eqref{eq:BeautyShorten2}, 
the excitation numbers of the highest submutiplet, \eqref{eq:BeautyOffsets2}, 
are
\[\label{eq:BeautyExcite2}
n_k=(0,1,2,4,2,1,0),\quad L=4.
\]
We therefore configure the simple roots as follows
\[\label{eq:BeautyTwistConfig}
k_j=(2,3,3,4,4,4,4,5,5,6).
\]
Now we note that the twist-two operators are unpaired states.
Therefore the configurations of Bethe roots
must be invariant under the symmetry $u_j\mapsto -u_j$ 
of the Bethe equations. This tells us
\[\label{eq:BeautyTwistSym}
u_1=u_{10}=0,\quad
u_2=-u_3,\quad
u_4=-u_5,\quad
u_6=-u_7,\quad
u_8=-u_9,
\]
which automatically satisfies 
the momentum constraint \eqref{eq:BeautyMomentum}.
Furthermore the excitations \eqref{eq:BeautyTwistExcite}
are invariant under flipping the Dynkin diagram,
$n_k\mapsto n_{8-k}$. This can be used to set
\[\label{eq:BeautyTwistSym2}
u_2=u_8.
\]
The Bethe equations \eqref{eq:BeautyBethe} are then solved exactly by 
\[\label{eq:BeautyTwistRoots}
u_2=\sqrt{\frac{5}{7}}\,,
\quad
u_{4,6}=\sqrt{\frac{65\pm 4\sqrt{205}}{140}}\,,
\]
which yields the energy \eqref{eq:BeautyEnergy}
\[\label{eq:BeautyTwistEngergy}
E=\frac{25}{3},\qquad \delta \Delta=\frac{\gym^2 N}{8\pi^2}\,E=\frac{25\gym^2 N}{24\pi^2}.
\]
This is the energy of the twist-two state at dimension four
\cite{Anselmi:1998ms}, see also \tabref{tab:Energies}.

\section{\ldots \& the Beast}
\label{sec:Beast}

The alert reader will have noticed that the Dynkin diagram 
in \eqref{eq:BeautyDynkin} is not the standard one found in
textbooks.
In fact, for superalgebras there are alternative 
choices for the Dynkin diagram. In this subsection
we will consider the `distinguished' Dynkin diagram
with one simple fermionic root
\[\label{eq:BeastDynkin}
\begin{minipage}{260pt}
\setlength{\unitlength}{1pt}%
\small\thicklines%
\begin{picture}(260,35)(-10,-10)
\put(  0,00){\circle{15}}%
\put(  0,10){\makebox(0,0)[b]{}}%
\put(  7,00){\line(1,0){26}}%
\put( 40,00){\circle{15}}%
\put( 40,15){\makebox(0,0)[b]{$-3$}}%
\put( 47,00){\line(1,0){26}}%
\put( 80,00){\circle{15}}%
\put( 80,15){\makebox(0,0)[b]{$+2$}}%
\put( 87,00){\line(1,0){26}}%
\put(120,00){\circle{15}}%
\put(120,15){\makebox(0,0)[b]{}}%
\put(115,-5){\line(1, 1){10}}%
\put(115, 5){\line(1,-1){10}}%
\put(127,00){\line(1,0){26}}%
\put(160,00){\circle{15}}%
\put(160,15){\makebox(0,0)[b]{}}%
\put(167,00){\line(1,0){26}}%
\put(200,00){\circle{15}}%
\put(200,15){\makebox(0,0)[b]{}}%
\put(207,00){\line(1,0){26}}%
\put(240,00){\circle{15}}%
\put(240,15){\makebox(0,0)[b]{}}%
\end{picture}
\end{minipage}
\]
Note, that for a different choice of Dynkin diagram, we get
different Dynkin labels for the spin representation.
What is more, the highest weight has actually changed as well
as we will see below.
The Cartan matrix and the representation vector read 
\[\label{eq:BeastCartan}
M=\left(\begin{array}{ccc|c|ccc}
+2&-1&  &  &  &  &   \\
-1&+2&-1&  &  &  &   \\
  &-1&+2&-1&  &  &   \\\hline
  &  &-1&  &+1&  &   \\\hline
  &  &  &+1&-2&+1&   \\
  &  &  &  &+1&-2&+1 \\
  &  &  &  &  &+1&-2
\end{array}\right),\qquad
V=\left(\begin{array}{r}
0\\-3\\2\\\hline0\\\hline0\\0\\0
\end{array}\right)
\]
The Bethe equations and momentum constraint are
still given by 
\eqref{eq:BeautyBethe},\eqref{eq:BeautyMomentum},
while the energy of a solution to the Bethe equations 
is now
\[\label{eq:BeastEnergy}
E=3L-\sum_{j=1}^n\lrbrk{\frac{i}{u_j+\sfrac{i}{2}V_{k_j}}-\frac{i}{u_j-\sfrac{i}{2}V_{k_j}}}.
\]
In this form of the Bethe equations, the vacuum has an energy,
it is a pseudovacuum and not the ground state of the theory.
The vacuum state is therefore \emph{not} the half-BPS state $\Tr Z^L$ in
\eqref{vac}, but instead
\[
\vac=\Tr \fldF^L
\]
i.e.~a composite operator of classical dimension $\Delta_0=2L$ 
and energy $E=3 L$ \cite{Beisert:2003jj} built from 
the field strength component $\fldF=\fldF_{1+i2,3+i4}$.
This vacuum configuration is the major difference 
to the Bethe ansatz discussed in the previous section.
What used to be a rather trivial state 
in one Bethe ansatz, becomes a highly 
excited state in the other. 
Nevertheless there is a duality between both sets of Bethe 
equations in that they must both yield the same spectra of energies.
This remarkable fact can be made use of in the usual sense of dualities.
To determine the energy of some set of states we apply that Bethe ansatz
which seems most suitable for the problem. For example,
the Bethe ansatz of the previous section is useful
for states with a low energy. 
In contrast, the Bethe ansatz discussed in this section 
seems most suitable for states
of a large chirality $B$ or 
of an energy around $3L$.

The number of excitations can be determined
as above. We note the action of the simple roots
in the oscillator picture of \cite{Beisert:2003jj}
\[\label{eq:BeastRoots}
\begin{minipage}{260pt}
\setlength{\unitlength}{1pt}%
\small\thicklines%
\begin{picture}(260,55)(-10,-30)
\put(  0,00){\circle{15}}%
\put(  0,15){\makebox(0,0)[b]{$n_1$}}%
\put(  0,-15){\makebox(0,0)[t]{$\oscb^\dagger_2 \oscb^1$}}%
\put(  7,00){\line(1,0){26}}%
\put( 40,00){\circle{15}}%
\put( 40,15){\makebox(0,0)[b]{$n_2$}}%
\put( 40,-15){\makebox(0,0)[t]{$\osca^\dagger_1 \oscb^\dagger_1$}}%
\put( 47,00){\line(1,0){26}}%
\put( 80,00){\circle{15}}%
\put( 80,15){\makebox(0,0)[b]{$n_3$}}%
\put( 80,-15){\makebox(0,0)[t]{$\osca^\dagger_2 \osca^1$}}%
\put( 87,00){\line(1,0){26}}%
\put(120,00){\circle{15}}%
\put(120,15){\makebox(0,0)[b]{$n_4$}}%
\put(120,-15){\makebox(0,0)[t]{$\oscc^\dagger_4 \osca^\dagger_2$}}%
\put(115,-5){\line(1, 1){10}}%
\put(115, 5){\line(1,-1){10}}%
\put(127,00){\line(1,0){26}}%
\put(160,00){\circle{15}}%
\put(160,15){\makebox(0,0)[b]{$n_5$}}%
\put(160,-15){\makebox(0,0)[t]{$\oscc^\dagger_3 \oscc^4$}}%
\put(167,00){\line(1,0){26}}%
\put(200,00){\circle{15}}%
\put(200,15){\makebox(0,0)[b]{$n_6$}}%
\put(200,-15){\makebox(0,0)[t]{$\oscc^\dagger_2 \oscc^3$}}%
\put(207,00){\line(1,0){26}}%
\put(240,00){\circle{15}}%
\put(240,15){\makebox(0,0)[b]{$n_7$}}%
\put(240,-15){\makebox(0,0)[t]{$\oscc^\dagger_1 \oscc^2$}}%
\end{picture}
\end{minipage}
\]
The vacuum corresponds to the configuration
$\osca^\dagger_1\osca^\dagger_1\vac$.
The numbers of excitations for a given weight are
\[\label{eq:BeastExcite}
n_k=\left(\begin{array}{l}
\sfrac{1}{2}(L-B)+\half\Delta_0-\half s_2-\phantom{2}L\\
\sfrac{2}{2}(L-B)+\phantom{\half}\Delta_0\phantom{\mathord{}-\half s_1}-2L\\
\sfrac{3}{2}(L-B)+\half\Delta_0-\half s_1\\
\sfrac{4}{2}(L-B)\\
\sfrac{3}{2}(L-B)-\half p-\quarter q_1-\sfrac{3}{4}q_2\\
\sfrac{2}{2}(L-B)-\phantom{\half}p-\half q_1-\half q_2\\
\sfrac{1}{2}(L-B)-\half p - \sfrac{3}{4}q_1 - \quarter q_2
\end{array}\right)\]
We note the bounds
\[\label{eq:BeastBounds}
0\leq n_1\leq n_2,
\quad
n_3\geq n_4\geq n_5\geq n_6\geq n_7\geq 0,
\]
as well as
\[\label{eq:BeastBounds2}
n_3-n_2\leq 2L,\quad
n_4-n_5,n_5-n_4,n_6-n_7,n_7\leq L.\]
We defer the discussion of highest weight
states and multiplet splitting 
in the context of the `distinguished' choice of 
Dynkin diagram to \appref{sec:Killing}.
There we also compare the
energies of all dimension four operators 
to the results of SYM.

\paragraph{Example.}

We see that the vacuum of this 
choice of Bethe equations has energy $3L$.
Clearly, this is not the ground state,
the ground state is half-BPS and has zero energy.
This situation is similar to the 
antiferromagnetic spin chain, where the ground state
is described by a highly excited state (in terms
of Bethe roots).
In this spin chain the first half-BPS 
multiplet with primary weight 
\[\label{eq:BeastBPSPrimary}
w_0=[2;0,0;0,2,0;0,2]
\]
has the highest weight (in the sense of highest chirality $B$)
\[\label{eq:BeastBPSHighest}
w=[4;0,0;0,0,0;2,2].
\]
This corresponds to the excitations and length
\[\label{eq:BeastBPSExcite}
n_k=(0,0,2,0,0,0,0),\quad k_j=(3,3),\quad L=2.\]
The Bethe equations and momentum constraint 
are solved exactly by
\[\label{eq:BeastBPSRoots}
u_{1,2}=\pm \frac{i}{\sqrt{3}}.
\]
Reassuringly, the energy due to the excitations
precisely cancels off the vacuum energy 
\[\label{eq:BeastBPSEngergy}
E=0.
\]
%

\paragraph{Spin Waves.}

For a low number of excitations but arbitrary length of the 
spin chain, the Bethe ansatz can quickly provide 
good approximations to the energies.
Especially in the case of two excitations 
it is often trivial to solve the Bethe equations exactly.
We will now discuss these spin wave solutions, which correspond
to small fluctuations of the vacuum. 
The major difference to the previous investigation of spin waves
in the context of $\superN=4$ SYM \cite{Minahan:2002ve}
is the modified vacuum. The solution is quite similar and
we will follow along the lines of \cite{Minahan:2002ve}.

Let us consider the following excitations of simple roots
\[\label{eq:BeastWavesExcite}
n_k=(0,1,1,0,0,0,0),\quad
k_j=(2,3).
\]
The momentum constraint implies
\[\label{eq:BeastWavesMom}
u_1=\sfrac{3}{2}u_2.
\]
and the two Bethe equations then collapse into
\[\label{eq:BeastWavesBethe}
\lrbrk{\frac{i+u_2}{i-u_2}}^{L-1}=-1.
\]
This equation has $L-1$ independent solutions 
\[\label{eq:BeastWavesRoots}
u_2=\cot\frac{\pi n}{L-1},\qquad
-\frac{L-1}{2}<n\leq \frac{L-1}{2}
\]
with energy
\[\label{eq:BeastWavesEnergy}
E=3L-\frac{2}{3}\sin^2\frac{\pi n}{L-1}.
\]
Note that the symmetry $u_j\mapsto -u_j$ translates into $n\mapsto -n$. 
The states $n$ and $-n$ therefore have degenerate energies and form a
pair unless $n=0$ or $n=(L-1)/2$.
Then the solutions $u_{1,2}=\infty$ or $u_{1,2}=0$, respectively
are invariant and do not pair up.

We can also consider the thermodynamic limit of a large $L$.
In leading order we can approximate the 
positions of the Bethe roots and the corresponding energy. 
An excitation of simple root 2 yields
\[\label{eq:BeastThermoRoot2}
u_{2,n}\approx \frac{3L}{2\pi n},\qquad
\delta E_{2,n}=\frac{4\pi^2 n^2}{3L^2}
\]
whereas for an excitation of simple root 3 we find
\[\label{eq:BeastThermoRoot3}
u_{3,n}\approx \frac{-2L}{2\pi n},\qquad
\delta E_{3,n}=-\frac{4\pi^2 n^2}{2L^2}.
\]
The sum of all mode numbers $n$ must vanish due to the momentum constraint.
Using this we quickly find an approximation to 
the above exactly solved spin wave.
\<\label{eq:BeastThermo23}
u_1\mathrel{}&\approx&\mathrel{}
\frac{3L}{2\pi n},\qquad
u_2 \approx
\frac{-2L}{2\pi (-n)},\nln
E\mathrel{}&\approx&\mathrel{} 3L+\frac{4\pi^2 n^2}{3L^2}-\frac{4\pi^2 n^2}{2L^2}=
3L-\frac{2\pi^2 n^2}{3L^2}.
\>
It would be interesting if some string theory solution describing 
this Bethe vacuum could be found. 
The spectrum of fluctuations, 
somewhat reminiscent of the plane wave spectrum, 
could then be compared.

\section{Outlook}
\label{sec:outlook}

We are confident that our proposal furnishes the tool for
very comprehensive ``spectroscopic'' studies of  
$\superN=4$ Yang-Mills theory. The recent progress with 
semi-classical approaches on the string side
\cite{Gubser:2002tv,Frolov:2002av,Russo:2002sr,Minahan:2002rc,
Tseytlin:2002ny,Frolov:2003tu,Frolov:2003xy} allows for very
interesting dynamical comparisons with gauge theory. In particular
our superchain and its associated Bethe equations should be
very useful for string motions that also involve the AdS part of the
background, or even joint motion in AdS and on the five sphere.
In all these applications string theory is able, in principle,
to predict anomalous dimensions for large dimension gauge theory
operators, to which the Bethe ansatz is ideally suited: 
On the level of the spin chain these situations correspond to the
thermodynamic limit of the chain, in which the Bethe equations
often become tractable \cite{Beisert:2003xu}.

It would be helpful to classify all possible ways to
write Bethe equations for the above $\alSL(4|4)$ super spin chain, and
identify the corresponding pseudovacua. Correspondingly one should
then find further spin wave solutions, describing the simplest fluctuations
around the pseudovacua. It would be interesting to see whether the
large dimension limits of the pseudovacuum states and their elementary
excitations play a special r\^ole in string theory.

One puzzling aspect of the observed integrable structures is
that they really correspond to ``hidden'' symmetries appearing
in large $N$ SYM$_4$. Clearly we would like to understand their
presence from the point of view of the gauge theory. Are they
a technical consequence of what one could call ``planar
supersymmetry'', or is there a deeper reason, possibly related
to string theory?

However, the most exciting open problem remains to identify the
integrable deformation of the super spin chain, with deformation parameter 
$\lambda=\gym^2 N$, that correctly
includes the effects of all quantum loops. 
This presumably would be a $\alPSL(4|4)$ super spin chain 
with long-range interactions.
A related question concerns the closure of the interacting
algebra and the consistency requirements imposed by it. 
How are these related to integrability and can they be used
to fix higher-loop corrections? 

\subsection*{Acknowledgements}

We would like to thank Gleb Arutyunov, Vladimir Dobrev, 
Stefano Kovacs, Joe Minahan, 
Jan Plefka and Kostya Zarembo for interesting discussions,
and A.~Belitsky, V.~Braun, A.~Gorsky and G.~Korchemsky
for email correspondence concerning the integrable structures
previously found in QCD.
N.B. dankt der \emph{Studien\-stiftung des Deutschen Volkes}
f\"ur die Unterst\"utzung durch ein Promotionsf\"orderungs\-stipendium.


\appendix

\section{Dissection of the Beast}
\label{sec:Killing}

In this appendix we present some more details on
the Bethe ansatz of \secref{sec:Beast}.
We will start by an investigation of highest weight states.
Next we will apply this to the Konishi state and its descendants,
for which we then solve the Bethe equations.
Finally, we present the set of all solutions
corresponding to states up to classical dimension four.

\subsection{Highest Weights}

A primary state is defined as a state
annihilated by the boosts $K,S,\dot S$.
This condition is identical to highest weight condition
in the sense of \secref{sec:Beauty}.
However, when we change the Dynkin diagram of the superalgebra,
we also change the notion of positive roots and highest weights.
For the `distinguished' Dynkin diagram of \secref{sec:Beast}
not all of the boosts $K,S,\dot S$ can be positive
roots simultaneously. 
Therefore primary weights
(where primary weight shall refer to highest weight in the
sense of \secref{sec:Beauty})
are not highest weights. Instead, a highest weight
state is annihilated by $K,S,\dot Q$.

For a given primary weight we need to find the corresponding 
highest weight. For a generic long multiplet with primary weight
$w$ this is achieved by adding 
\[\label{eq:BeastOffsetLong}
\delta w=[4;0,0;0,0,0;4,0].
\]
It corresponds to acting with $\dot Q^8$ on the primary state.
A further application of $\dot Q$,
as well as $K,S$ which commute with $\dot Q^8$, will therefore kill the state.

In the classical $\alSL(4|4)$ there are various 
shortening conditions that affect the highest weight.
For a half-BPS state we add
\[\label{eq:BeastOffsetHalf}
\delta w_{1/2}=[2;0,0;0,-2;0;2,0]
\]
instead.
The fundamental multiplet $V\indup{F}$ with primary weight 
$[1;0,0;0,1,0;0,1]$ is subject to an additional shortening condition. 
The highest weight is $[2;2,0;0,0,0;1,1]$ and corresponds to the field
$\fldF_{1+i2,3+i4}$ of $\superN=4$. The vacuum is constructed
from exactly this field.

There are two shortening conditions for unprotected multiplets,
see \eqref{eq:BeautyShorten}.
When shortening condition I applies, the multiplet
splits in two with highest weights
\[\label{eq:BeastOffsetI}
w+\delta w+\delta w\indup{I},\qquad
w+\delta w\indup{I+}.\]
where
\<\label{eq:BeastOffsetI2}
\delta w\indup{I}\eq[0;0,0;0,0,0;-1,+1],
\nln
\delta w\indup{I+}\eq
\left\{\begin{array}{l}
[+3.5;+1,0;-1,\phantom{+}0,\phantom{+}0;+3.5,0],\\{}
[+3.0;+2,0;\phantom{+}0,-1,\phantom{+}0;+3.0,0],\\{}
[+2.5;+3,0;\phantom{+}0,\phantom{+}0,-1;+2.5,0],\\{}
[+2.0;+4,0;\phantom{+}0,\phantom{+}0,\phantom{+}0;+2.0,0].
\end{array}\right.
\>
For $\delta w\indup{I+}$ it is understood that the topmost line 
applies for which the resulting labels are positive.
When shortening condition II applies, the highest weights are
\[\label{eq:BeastOffsetII}
w+\delta w,\qquad
w+\delta w\indup{II}+\delta w.
\]
with 
\[\label{eq:BeastOffsetII2}
\delta w\indup{II}=
\arraycolsep0pt
\left\{\begin{array}{l}
[+0.5;0,-1;0,0,+1;+0.5,+1],\\{}
[+1.0;0,\phantom{+}0;0,0,+2;\phantom{+}0.0,+1].
\end{array}\right.
\]
Finally, if both shortening conditions apply, the four highest weights are
\[\label{eq:BeastOffsetIII}
\begin{array}{l}
w+\delta w+\delta w\indup{I},\\{}
w+\delta w\indup{I+},\\{}
w+\delta w\indup{II}+\delta w+\delta w\indup{I},\\{}
w+\delta w\indup{II}+\delta w\indup{I+}.
\end{array}\]
Note that to determine the last highest weight, it is necessary to
choose the correct $\delta w\indup{I+}$ for 
$w+\delta w\indup{II}$.
The submultiplets join in the interacting theory.
In each of the three cases
\eqref{eq:BeastOffsetI},\eqref{eq:BeastOffsetII},\eqref{eq:BeastOffsetIII}
we have listed the highest weight of the long multiplet first.

\subsection{Konishi Submultiplets}

The Konishi state has primary weight
\[
w_0=[2;0,0;0,0,0;0,2].
\]
This is at both unitarity bounds.
According to the rules of the previous section 
we find the highest weights of the four submultiplets
\<
w_1\eq [4.0;4,0;0,0,0;2.0,2],
\nln
w_2\eq [6.0;0,0;0,0,0;3.0,3],
\nln
w_3\eq [5.5;3,0;0,0,1,2.5,3],
\nln
w_4\eq [7.0;0,0;0,0,2;3.0,4].
\>
The excitation numbers are determined by \eqref{eq:BeastExcite}
\<
n_{1,k}\eq (0,0,0,0,0,0,0),\quad L=2,
\nln
n_{2,k}\eq (0,0,3,0,0,0,0),\quad L=3,
\nln
n_{3,k}\eq (0,0,2,1,0,0,0),\quad L=3,
\nln
n_{4,k}\eq (0,0,5,2,0,0,0),\quad L=4.
\>
The first configuration is the vacuum, 
we find the energy $E=3L=6$ straight away.

For the second configuration with $k_j=(3,3,3)$ we note that the
roots must be invariant under the map $u_j\mapsto -u_j$ 
because the Konishi state is unpaired.
Therefore one root is zero while the other two sum to zero.
The only way to satisfy the momentum constraint is to 
pick the singular configuration
\[\label{eq:BeastKoni2}
u_{1,2}=\pm i,\quad u_3=0.
\]
This has to be regularized in the usual way,
we again find the energy $E=6$.
As an aside, this state 
has the highest weight for the interacting
multiplet. At leading order, 
it corresponds to the operators 
$\varepsilon^{\beta\gamma}
\varepsilon^{\delta\varepsilon}
\varepsilon^{\eta\alpha}
\Tr \fldF_{\alpha\beta}\fldF_{\gamma\delta}\fldF_{\varepsilon\eta}$
of $\superN=4$ SYM.

For the third configuration with $k_j=(3,3,4)$
it is straightforward to find and verify
\[\label{eq:BeastKoni3}
u_{1,2}=\pm\frac{1}{\sqrt{3}},\quad u_3=0
\]
as a solution to the Bethe equations and momentum constraint
with energy $E=6$.

For the last configuration with $k_j=(3,3,3,3,3,4,4)$
we again require a singular solution.
Taking the roots \eqref{eq:BeastKoni2} as an ansatz, 
the remaining roots can be found
\[\label{eq:BeastKoni4}
u_{1,2}=\pm i,\quad u_3=0,\quad
u_{4,5}=\pm \frac{1}{\sqrt{3}},\quad
u_{6,7}=\pm \sqrt{\frac{7}{20}}.
\]
They satisfy the momentum constraint 
and yield, not surprisingly, $E=6$.

We have seen that all submultiplet of the Konishi multiplet 
have the same energy. This is somewhat remarkable, 
as they have rather different configurations of roots. 

\subsection{Bethe Roots for Dimension Four States}

Here we will present the Bethe roots
for operators up to dimensions four.
This demonstrates that also the `distinguished' 
choice of Dynkin diagram with a pseudovacuum
leads to the correct results.
We present the complete list of operators and anomalous dimensions
in \tabref{tab:Energies}.
The Bethe roots which solve the Bethe equations and momentum 
constraint are given below. The energies agree with 
\tabref{tab:Energies}, we do not list them again.
\begin{table}\centering
$\begin{array}{|c|cccc|l|}\hline
\Delta_0&\alSL(2)^2&\alSL(4)&B&L&E^P\\\hline
2&[0,0]&[0,2,0]&0&2&0^+ \\
 &[0,0]&[0,0,0]&0&2&6^+ \\
\hline
3&[0,0]&[0,3,0]&0&3&0^- \\
 &[0,0]&[0,1,0]&0&3&4^- \\
\hline
4&[0,0]&[0,4,0]&0&4&0^+ \\
 &[0,0]&[0,2,0]&0&4&(5\pm \sqrt{5})^+ \\
 &[0,0]&[1,0,1]&0&4&6^- \\
 &[0,0]&[0,0,0]&0&4&\frac{1}{2}(13\pm \sqrt{41})^+ \\
 &[2,0]&[0,0,0]&1&3&9^- \hfill+\mathord{\mbox{conj.}}\\
 &[1,1]&[0,1,0]&0&3&\frac{15}{2}^{\pm}\\
 &[2,2]&[0,0,0]&0&2&\frac{25}{3}^{+}\\
\hline
\end{array}$
\caption{All energies $E$ of primary states with $\Delta_0\leq 4$,
see \cite{Beisert:2003jj}.
The parity $P$ is related to parity under inversion of the spin chain,
parity $\pm$ indicates a degenerate pair.
The label `+conj.' indicates conjugate states with
$\alSL(2)^2,\alSL(4)$ labels reversed and opposite chirality $B$.}
\label{tab:Energies}
\end{table}

\paragraph{[2;0,0;0,2,0;0,2].} 
$k_j=(3,3)$, $L=2$. See \eqref{eq:BeastBPSRoots}.
\[
u_{1,2}=\pm \frac{i}{\sqrt{3}}\,.
\]

\paragraph{[2;0,0;0,0,0;0,2].} 
$k_j=()$, $L=2$. Trivial.

\paragraph{[3;0,0;0,3,0;0,3].} 
$k_j=(3,3,3,3,4,4,5)$, $L=3$. 
\[
u_{1,2,3,4}=\pm\sqrt{\frac{-3\pm i\sqrt{15}}{12}}\,,\quad
u_{5,6}=i\sqrt{\frac{5}{12}}\,,\quad
u_7=0.
\]

\paragraph{[3;0,0;0,1,0;0,3].} 
$k_j=(3,3)$, $L=3$.
\[
u_{1,2}=\pm \frac{i}{\sqrt{5}}\,.
\]

\paragraph{[4;0,0;0,4,0;0,4].} 
$k_j=(3,3,3,3,3,3,4,4,4,4,5,5)$, $L=4$.
This is left as an exercise for the reader. Good luck!

\paragraph{[4;0,0;0,2,0;0,4].} 
$k_j=(3,3,3,3,4,4,5)$, $L=4$. Two states distinguished by $\pm'$.
\[
u_{1,2,3,4}=\pm\sqrt{\frac{\mathbin{\pm'}\sqrt{5}\pm i(5\mathbin{\pm'}3\sqrt{5})}{15}}\,,\quad
u_{5,6}=\pm \sqrt{\frac{-15\mathbin{\mp'}2\sqrt{5}}{36}}\,,\quad
u_7=0.
\]

\paragraph{[4;0,0;1,0,1;0,4].} 
$k_j=(3,3,3,3,4)$, $L=4$. 
\[
u_{1,2,3,4}=\frac{\pm 1\pm i}{\sqrt{2\sqrt{3}}}\,,\quad
u_{5}=0.
\]

\paragraph{[4;0,0;0,0,0;0,4].} Two states distinguished by $\pm'$.
$k_j=(3,3,3,3)$, $L=4$. 
\[
u_{1,2,3,4}=\pm \sqrt{\frac{-9\mathbin{\mp'}\sqrt{41}}{12}\pm \sqrt{\frac{173\mathbin{\pm'}33\sqrt{41}}{360}}}\,.
\]

\paragraph{[4;2,0;0,0,0;1,3].} 
$k_j=()$, $L=3$. Trivial.

\paragraph{[4;0,2;0,0,0;$-1$,3].} 
$k_j=(2,2,3,3,3,3)$, $L=3$. 
\[
u_{1,2}=\pm \sqrt{\frac{17}{60}}\,,\quad
u_{3,4,5,6}=\pm \sqrt{\frac{-39\pm\sqrt{993}}{30}}\,.
\]

\paragraph{[4;1,1;0,1,0;0,3].} 
$k_j=(2,3,3)$, $L=3$. A pair of states distinguished by $\pm'$.
\[
u_{1}=\pm'\sqrt{\frac{5}{28}}\,,\quad
u_{2,3}=\mp'\sqrt{\frac{1\pm i\sqrt{399}}{30}}\,.
\]

\paragraph{[4;2,2;0,0,0;0,2].} 
$k_j=(2,2)$, $L=2$. 
\[
u_{1,2}=\pm\sqrt{\frac{9}{28}}\,.
\]
%


\bibliography{beauty}
\bibliographystyle{nb}

\end{document}